\\
Title: Stability of the Vacuum Non Singular Black Hole
Author:Gamal G. L. Nashed
Journal-ref: Chaos, Solitons & Fractals 15 (2003), 841.
\\
The singularity of the black hole solutions obtained  before in
M\o ller's theory are studied. It is found that although the two
solutions reproduce the same associated metric the asymptotic
behavior of the scalars of torsion tensor and basic vector are
quite different. The stability of the associated metric of those
solutions which is spherically symmetric non singular black hole
is studied using the equations of geodesic deviation. The
condition for the stability is obtained. From this condition the
stability of the Schwarzschild solution and di Sitter solution can
be obtained.
\\
\documentstyle[12pt]{article}
\newlength{\extraspace}
\newlength{\extraspaces}
\newcommand{\be}{\begin{equation}
\addtolength{\abovedisplayskip}{\extraspaces}
\addtolength{\belowdisplayskip}{\extraspaces}
\addtolength{\abovedisplayshortskip}{\extraspace}
\addtolength{\belowdisplayshortskip}{\extraspace}}
\newcommand{\ee}{\end{equation}}

\newcommand{\ba}{\begin{eqnarray}
\addtolength{\abovedisplayskip}{\extraspaces}
\addtolength{\belowdisplayskip}{\extraspaces}
\addtolength{\abovedisplayshortskip}{\extraspace}
\addtolength{\belowdisplayshortskip}{\extraspace}}
\newcommand{\ea}{\end{eqnarray}}

\newcommand{\nonu}{\nonumber \\[.5mm]}
\newcommand{\A}{&\!\!\!}

\newcommand{\il}{\lambda_{{}_{{}_{\!\!\!\!\scriptstyle{i}}}}}

\newcommand{\bl}{\lambda_{{}_{{}_{\!\!\!\!\scriptstyle{0}}}}}

\newcommand{\newsection}[1]{
\vspace{7mm} \pagebreak[3] \addtocounter{section}{1}
\setcounter{subsection}{0} \setcounter{footnote}{0}
\begin{center}
{\large {\bf \thesection. #1}}
\end{center}
\nopagebreak
\medskip
\nopagebreak \hspace{3mm}}

\begin{document}

\begin{center}
{{\bf Stability of the Vacuum Non Singular Black Hole}}
\end{center}
\centerline{ Gamal G.L. Nashed}

\bigskip

\centerline{{\it Mathematics Department, Faculty of Science, Ain
Shams University, Cairo, Egypt }}

\bigskip
 \centerline{ e-mail:nashed@asunet.shams.edu.eg}

\hspace{2cm}
\\
\\
\\
\\
\\
\\
\\
\\

The singularity of the black hole solutions obtained  before in
M\o ller's theory are studied. It is found that although the two
solutions reproduce the same associated metric the asymptotic
behavior of the scalars of torsion tensor and basic vector are
quite different. The stability of the associated metric of those
solutions which is spherically symmetric non singular black hole
is studied using the equations of geodesic deviation. The
condition for the stability is obtained. From this condition the
stability of the Schwarzschild solution and di Sitter solution can
be obtained.
\newpage

\begin{center}
\newsection{\bf Introduction}
\end{center}

A black holes is a singularity in Einstein field equations. This
can be happen when a gravitational collapse takes place and
continue until the surface of the star approaches the
Schwarzschild radius, i.e., $r=2m$ \cite{Di5}. Hawking and
collaborators discovered that the laws of thermodynamics have an
exact analogues in the properties of black holes \cite{Di5,
Hw1,Hw2,Hw3}. As a black hole emits particles, its mass and size
steadily decrease. This makes it easier to tunnel out and so the
emission will continue at an ever-increasing rate until eventually
the black hole radiates itself out of existence. In the long run,
every black hole in the universe will evaporates in this way.
Sidharth \cite{Bg} shown that a particle
 can be treated as a relativistic vortex, that is a vortex where the
 velocity of a circulation equals that of light or a spherical shell,
 whose constituents are again rotating with the velocity of light
 or as a black hole described by the Kerr-Newman metric for a spin
 $\displaystyle{1 \over 2}$  particles.

Dymnikova \cite{Di} derived a static spherically symmetric  black
 hole solution in orthodox general relativity
 assuming a specific
form of the stress-energy momentum tensor. This solution
practically coincides with the Schwarzschild solution for large
$r$,  for small $r$ it behaves like the de Sitter solution and
describes a spherically symmetric black hole singularity free
everywhere \cite{Di}. It is shown that the metric of this solution
has two event horizons, one related to the external horizon and
the other is the internal horizon which is the Cauchy horizon
\cite{Di}. It is shown that both of the horizons are removable.
Dymnikova \cite{Di} has shown that this solution is regular at
$r=0$ and so it is nonsingular everywhere, so it is called a
"nonsingular black hole". It has been proved that it is possible
to treat this specific form of the stress-energy momentum tensor
as corresponding to an r-dependent cosmological term $\Lambda_{\mu
\nu}$, varying from $\Lambda_{\mu \nu}=\Lambda g_{\mu \nu}$ as  $r
\rightarrow 0$ to $\Lambda_{\mu \nu}=0$ as $r \rightarrow \infty$
\cite{Di1}. More recently \cite{Di2}, the spherically symmetric
nonsingular black hole has been used to
 prove that a baby universe inside a $\Lambda$ black hole can be
obtained in the case of an eternal black hole. Also it has been
shown that the probability of a quantum birth of a baby universe
can not be neglected due to the existence of an infinite number of
$\Lambda$ white hole structures.

In an earlier paper \cite{Ga} the author used a spherically
symmetric tetrad constructed by Robertson \cite{Ro} to derive two
different spherically symmetric vacuum nonsingular black hole
solutions of M\o ller's field equations assuming the same form of
the vacuum stress-energy momentum tensor given in \cite{Di}. He
also calculated the energy content of these solutions.

It is the aim of the present paper to study the singularity of the
two solutions obtained before \cite{Ga} and then derive the
condition for the stability of these solutions using the geodesic
deviation \cite{WB}. In section 2, a brief review of the two
solutions obtained before are given. The singularity problem of
these solutions are studied in section 3. The condition of the
stability for the vacuum non singular black hole solution is given
in section 4. Section 5 is devoted to main results.
\\
\\

Computer algebra system Maple 6 is used in some calculations.

\begin{center}
\newsection{\bf Spherically symmetric nonsingular black hole solutions}
\end{center}

Dymnikova \cite{Di} in 1992 has obtained a static spherically
symmetric nonsingular black hole solution in general relativity
which is expressed by \be ds^2=(1-{R_g(r) \over r})
dt^2-\displaystyle{dr^2 \over (1-\displaystyle{R_g(r) \over r})}
-r^2 d\theta^2-r^2 \sin^2\theta d\phi^2, \ee where \be
R_g(r)=r_g(1-e^{-r^3/{r_1}^3}), \qquad {r_1}^3={r_0}^2r_g, \qquad
{r_0}^2={3 \over 8\pi \epsilon_0}, \qquad and \qquad r_g=2M. \ee
This black hole is regular at r=0 as we will see from the study of
the singularity problem. For the Einstein field equation to be
satisfied the stress-energy momentum tensor must take the form
\cite{Di} \ba
{T_0}^0={T_1}^1 \A=\A \epsilon_0 e^{-r^3/{r_1}^3},\nonu %
{T_2}^2={T_3}^3 \A=\A \epsilon_0 e^{-r^3/{r_1}^3} \left(1-{3r^3
\over 2{r_1}^3} \right).
 \ea

In a previous paper the author used the teleparallel spacetime in
which the fundamental fields of gravitation are the parallel
vector fields ${b_k}^\mu$. The component of the metric tensor
$g_{\mu \nu}$ are related to the dual components ${b^k}_\mu$ of
the parallel vector fields by the relation \be g_{\mu \nu}=\eta_{i
j} {b^i}_\mu{b^j}_\nu, \ee
 where $\eta_{i j}=diag.(-,+,+,+)$. The nonsymmetric connection
 ${\Gamma^\lambda}_{\mu \nu}$\footnote{Latin indices $(i,j,k,\cdots)$ designate the vector
number, which runs from $(0)$ to
 $(3)$, while Greek indices $(\mu,\nu,\rho, \cdots)$ designate the world-vector components
running from 0 to 3. The spatial part of Latin indices is denoted
by $(a,b,c,\cdots)$, while that of Greek indices by $(\alpha,
\beta,\gamma,\cdots)$.} are defined by
 \be
{\Gamma^\lambda}_{\mu \nu} ={b_k}^\lambda {b^k}_{\mu,\nu}, \ee as
a result of the absolute parallelism \cite{MW}.

The gravitational Lagrangian ${\it L}$ of this theory is an
invariant constructed from the quadratic terms of the torsion
tensor
 \be
 {T^\lambda}_{\mu \nu} \stackrel{\rm def.}{=}{\Gamma^\lambda}_{\mu
 \nu}-{\Gamma^\lambda}_{\nu \mu}. \ee
  The gravitational Lagrangian is given by \cite{HS}
\be {\cal L} \stackrel{\rm def.}{=} -{1 \over 3\kappa} \left(
t^{\mu \nu \lambda} t_{\mu \nu \lambda}- v^\mu v_\mu \right)+\xi
a^\mu a_\mu. \ee  Here $\xi$ is a constant parameter, $\kappa$ is
the Einstein gravitational constant and $t_{\mu \nu \lambda},
v_\mu$ and $a_\mu$ are the irreducible components of the torsion
tensor: \ba t_{\lambda \mu \nu}\A=\A {1 \over 2} \left(T_{\lambda
\mu \nu}+T_{\mu \lambda \nu} \right) +{1 \over 6} \left( g_{\nu
\lambda}V_\mu+g_{\mu \nu}V_\lambda \right) -{1 \over 3} g_{\lambda
\mu}V_\nu,\nonu
V_\mu \A=\A {T^\lambda}_{\lambda \nu}, \nonu
a_\mu \A=\A {1 \over 6} \epsilon_{\mu \nu \rho \sigma}T^{\nu \rho
\sigma},
 \ea
where $\epsilon_{\mu \nu \rho \sigma}$ is defined by \be
\epsilon_{\mu \nu \rho \sigma} \stackrel{\rm def.}{=} \sqrt{-g}
 \delta_{\mu \nu \rho \sigma} \ee with $\delta_{\mu \nu \rho
\sigma}$ being completely antisymmetric and normalized as
$\delta_{0123}=-1$.

By applying the variational principle to the Lagrangian (7), the
gravitational field equation are given by \cite{HS}\footnote{We
will denote the symmetric part by ( \ ), for example, $A_{(\mu
\nu)}=(1/2)( A_{\mu \nu}+A_{\nu \mu})$ and the  antisymmetric part
by the square bracket [\ ], $A_{[\mu \nu]}=(1/2)( A_{\mu
\nu}-A_{\nu \mu})$ .}: \be G_{\mu \nu} +K_{\mu \nu} = -{\kappa}
T_{(\mu \nu)}, \ee \be {b^i}_\mu{b^j}_\nu
\partial_\lambda(\sqrt{-g} {J_{i j}}^{\lambda})=\lambda
\sqrt{-g}T_{[\mu \nu]}, \ee where the Einstein tensor $G_{\mu
\nu}$ is defined by \be G_{\mu \nu}=R_{\mu \nu}-{1 \over 2} g_{\mu
\nu} R, \ee \be {R^\rho}_{\sigma \mu \nu}=\partial_\nu \left
\{_{\sigma \mu}^\rho \right\}-
\partial_\mu \left \{_{\sigma \nu}^\rho \right\} +
\left \{_{\sigma \mu}^\tau \right\} \left\{_{\tau \nu}^\rho
\right\}
 - \left \{_{\tau \mu}^\rho \right\} \left \{_{\sigma
\nu}^\tau \right\}, \ee \be R_{\mu \nu}={R^\rho}_{\mu \rho \nu},
\ee \be R=g^{\mu \nu}R_{\mu \nu}, \ee and $\left \{_{\mu
\nu}^\lambda \right\}$ is the Christoffel second kind define by
\be \left \{_{\mu \nu}^\lambda \right\}={1 \over 2} g^{\sigma
\lambda} \left(g_{\mu \sigma,\ \nu}+g_{\nu \sigma,\ \mu}-g_{\mu
\nu, \ \sigma} \right), \ee where $,$ is the differentiation with
respect to the coordinate
 and $T_{\mu \nu}$ is the energy-momentum
tensor of a source field of the Lagrangian $L_m$ \be T^{\mu \nu} =
{1 \over \sqrt{-g}} {\delta {\cal L}_M \over \delta {b^k}_\nu}
b^{k \mu} \ee with ${L_M}= {{\cal L}_M/\sqrt{-g}}$. The tensors
$K_{\mu \nu}$ and $J_{i j \mu}$ are defined as \be K_{\mu
\nu}={\kappa \over \lambda}\left( {1 \over 2}
\left[{\epsilon_\mu}^{\rho \sigma \lambda}(T_{\nu \rho
\sigma}-T_{\rho \sigma \nu})+{\epsilon_\nu}^{\rho \sigma
\lambda}(T_{\mu \rho \sigma}-T_{\rho \sigma \mu})
\right]a_\lambda-{3 \over 2} a_\mu a_\nu-{3 \over 4}g_{\mu \nu}
a^\lambda a_\lambda \right), \ee \be J_{i j \mu}=-{3 \over 2}
{b_i}^\rho {b_j}^\sigma \epsilon_{\rho \sigma \mu \nu} a^\nu, \ee
respectively. The dimensionless parameter $\lambda$ is defined by
\be {1 \over \lambda}={4 \over 9} \xi+{1 \over 3 \kappa}.\ee

The structure of the Weintzenb${\ddot o}$ck spaces with spherical
symmetry and has three unknown functions of radial coordinate is
given by Robertson \cite{Ro}  in the form
 \be
\left(\il^\mu \right)= \left( \matrix{ iA & iDr & 0 & 0
\vspace{3mm} \cr 0 & B \sin\theta \cos\phi & \displaystyle{B \over
r}\cos\theta \cos\phi
 & -\displaystyle{B \sin\phi \over r \sin\theta} \vspace{3mm} \cr
0 & B \sin\theta \sin\phi & \displaystyle{B \over r}\cos\theta
\sin\phi
 & \displaystyle{B \cos\phi \over r \sin\theta} \vspace{3mm} \cr
0 & B \cos\theta & -\displaystyle{B \over r}\sin\theta  & 0 \cr }
\right), \ee where the vector $\bl^\mu$ has taken to be imaginary
in order to preserve the Lorentz signature for the metric, i.e,
the functions $A$ and $D$ have to be taken as imaginary. Applying
(21) to (10) and (11) the author got \ba \A \A when  \qquad A=1,
\qquad B=1 \nonu
\A \A D=\sqrt{{2m\over r3}\left(1-e^{-r^3/{r_1}^3}\right)}, \ea
and when $D=0$ \ba A \A=\A {1 \over \sqrt{1- \displaystyle{2m
\over R}\left(1-e^{-R^3/{r_1}^3}\right)}}, \nonu
B \A=\A \sqrt{1-\displaystyle{2m \over
R}\left(1-e^{-R^3/{r_1}^3}\right)}. \ea Using (4) the associated
metric of the two solutions (22) and (23) is found to be the same
as (1) with the stress-energy momentum tensor given by (2). Now we
are going to study the singularity problem for both (22) and (23).
\newsection{Singularities}
In teleparallel theories we mean by singularity of space-time
\cite{KT} the singularity of the scalar concomitants of the
torsion and curvature tensors.

Using (13), (14), (15), (6) and (8) the scalars of the
Riemann-Christofell curvature tensor, Ricci tensor, Ricci scalar,
torsion tensor, basic vector, traceless part and the axial vector
part of the space-time given by the solution (22) are given by
\newpage
 \ba R^{\mu \nu \lambda \sigma}R_{\mu \nu \lambda
\sigma} \A = \A {12m^2 \over {r_1}^{12}r^6}
\Biggl[(24{r_1}^6r^6+8{r_1}^9r^3+4{r_1}^{12}+27r^{12})e^{-2r^3/{r_1}^{3}}
\nonu
\A \A
-4(3{r_1}^6r^6+2{r_1}^9r^3+2{r_1}^{12})e^{-r^3/{r_1}^{3}}+4{r_1}^{12}
\Biggr], \nonu
R^{\mu \nu}R_{\mu \nu} \A=\A  {18m^2 \over {r_1}^{12}}
\left[(8{r_1}^6-12{r_1}^3r^3+9r^{6}) \right] e^{-2r^3/{r_1}^{3}}
\nonu
R \A=\A {6m \over {r_1}^{6}} \left[(3r^3-4{r_1}^3) \right]
e^{-r^3/{r_1}^{3}} \nonu
T^{\mu \nu \lambda}T_{\mu \nu \lambda} \A=\A {3m \over
{r_1}^{6}r^3\left( e^{-r^3/{r_1}^{3}}-1 \right)}
\Biggl[(2{r_1}^3r^3+3{r_1}^6r^3+3r^{6})e^{-2r^3/{r_1}^{3}} \nonu
\A \A -(2{r_1}^3r^3+6{r_1}^6)e^{-r^3/{r_1}^{3}}+3{r_1}^{6}
\Biggr], \nonu
V^\mu V_\mu \A=\A {9m \over 2{r_1}^{6}r^3\left(
e^{-r^3/{r_1}^{3}}-1 \right)} \left[{r_1}^3-({r_1}^3-r^3)
e^{-2r^3/{r_1}^{3}} \right] \nonu
t^{\mu \nu \lambda}t_{\mu \nu \lambda} \A=\A {9m \over
2{r_1}^{6}r^3\left( 1-e^{-r^3/{r_1}^{3}} \right)}
\left[{r_1}^3-({r_1}^3-r^3) e^{-2r^3/{r_1}^{3}} \right]^2 \nonu
a^\mu a_\mu\A =\A 0.
 \ea
The scalars of the Riemann-Christofell curvature tensor, Ricci
tensor and  Ricci scalar of the solution (23) are the same as
given by (24). This is a logic results since both the solutions
reproduce the same metric tensor and these scalars mainly depend
on the metric tensor. The scalars of torsion tensor, basic vector,
traceless part and the axial vector part of the  space-time given
by the solution (23) are given by \ba T^{\mu \nu \lambda}T_{\mu
\nu \lambda} \A=\A {2 \over {r_1}^{6}r^3\left(r-2m+
2me^{-r^3/{r_1}^{3}} \right)} \Biggl[{r_1}^{6}\left(2r-3m+
3me^{-r^3/{r_1}^{3}} \right)^2\nonu
\A \A - 4\sqrt{r}{r_1}^{6}\left(r-2m+ 2me^{-r^3/{r_1}^{3}}
\right)^{3/2} +3m^2r^3e^{-r^3/{r_1}^{3}}
\left(3r^3e^{-r^3/{r_1}^{3}}-2{r_1}^3+2{r_1}^3e^{-r^3/{r_1}^{3}}
\right) \Biggr] \nonu
V^\mu V_\mu \A=\A  {1 \over {r_1}^{6}r^4\left(r-2m+
2me^{-r^3/{r_1}^{3}} \right)^2} \Biggl[ 2r {r_1}^3 \left(r-2m+
2me^{-r^3/{r_1}^{3}} \right)\nonu
\A \A +\sqrt{r2-2mr+ 2mr e^{-r^3/{r_1}^{3}}} \left \{ 3{r_1}^{3}
m-3{r_1}^{3}m
e^{-r^3/{r_1}^{3}}-2r{r_1}^{3}+3mr^3e^{-r^3/{r_1}^{3}} \right \}
\Biggr]^2, \nonu
t^{\mu \nu \lambda}t_{\mu \nu \lambda} \A=\A {1 \over
{r_1}^{6}r^4\left(r-2m+ 2me^{-r^3/{r_1}^{3}} \right)^2} \Biggl[ r
{r_1}^3 \left(r-2m+ 2me^{-r^3/{r_1}^{3}} \right)\nonu
\A \A +\sqrt{r2-2mr+ 2mr e^{-r^3/{r_1}^{3}}} \left \{ 3{r_1}^{3}
m-3{r_1}^{3}m
e^{-r^3/{r_1}^{3}}-r{r_1}^{3}-3mr^3e^{-r^3/{r_1}^{3}} \right \}
\Biggr]^2.
 \ea
 As is clear  from (24) and (25) that the scalars of the torsion, basic
 vector and the traceless part of the two solutions (22) and (23)
 are quite different in spite that they gave the same associated metric
 (1). Also as we see from (24) and (25) that as $r \rightarrow 0$ all
 the scalars take finite value.
 \newsection{The Stability condition}
 In the background of gravitational field the trajectories are
 represented by the geodesic equation
 \be
 {d^2 x^\lambda \over ds^2}+ \left\{^\lambda_{ \mu \nu} \right \}
 {d x^\mu \over ds}{d x^\nu \over ds}=0,
 \ee
 where $\displaystyle{d x^\mu \over ds}$ is the velocity four vector, s is a parameter varying
 along the geodesic. It is well know that the perturbation of the
 geodesic will lead to deviation \cite{Di5}
 \be
 {d^2 \zeta^\lambda \over ds^2}+ 2\left\{^\lambda_{ \mu \nu} \right \}
 {d x^\mu \over ds}{d \zeta^\nu \over ds}+
 \left\{^\lambda_{ \mu \nu} \right \}_{,\ \rho}
 {d x^\mu \over ds}{d x^\nu \over ds}\zeta^\rho=0,
 \ee
where $\zeta^\rho$ is the deviation 4-vector.

Applying (26), (27)  in (1) we get for the geodesic equations \be
{d^2 t \over ds^2}=0, \qquad {1 \over 2} \eta'(r)\left({d t \over
ds}\right)^2-r\left({d \phi \over ds}\right)^2=0, \qquad {d^2
\theta \over ds^2}=0,\qquad {d^2 \phi \over ds^2}=0,\ee and for
the geodesic deviation \ba \A \A {d^2 \zeta^0 \over
ds^2}+{\eta'(r) \over \eta(r)}{dt \over ds}{d \zeta^1 \over ds}=0
\nonu
\A \A {d^2 \zeta^1 \over ds^2}+\eta(r)\eta'(r) {dt \over ds}{d
\zeta^0 \over ds}-2r \eta(r) {d \phi \over ds}{d \zeta^3 \over
ds}\nonu
\A \A +\left[{1 \over 2}\left(\eta'^2(r)+\eta(r) \eta''(r)
\right)\left({dt \over ds}\right)^2-\left(\eta(r)+r\eta'(r)
\right) \left({d\phi \over ds}\right)^2 \right]\zeta^1=0, \nonu
\A \A {d^2 \zeta^2 \over ds^2}+\left({d\phi \over ds}\right)^2
\zeta^2=0, \nonu
\A \A {d^2 \zeta^3 \over ds^2}+{2 \over r}{d\phi \over ds} {d
\zeta^1 \over ds}=0, \ea where $\eta(r)=(1-\displaystyle{R_g(r)
\over r})$, $\eta'(r)=\displaystyle{d\eta(r) \over dr}$ and we
have consider the circular orbit in the plane \be \theta={\pi
\over 2}, \qquad {d\theta \over ds}=0, \qquad  {d r \over ds}=0.
\ee Using (30) in (1) we get \be \eta(r)\left({dt \over
ds}\right)^2-r^2\left({d\phi \over ds}\right)^2=1, \ee from (28)
and (31) we obtain \be \left({d\phi \over ds}\right)^2={\eta'(r)
\over r(2\eta(r)-r\eta'(r))}, \qquad \left({dt \over
ds}\right)^2={2 \over 2\eta(r)-r\eta'(r)}. \ee

The variable $s$ in (29) can be eliminated and we can rewrite it
in the form \ba \A \A {d^2 \zeta^0 \over d\phi^2}+{\eta'(r) \over
\eta(r)}{dt \over d\phi}{d \zeta^1 \over d\phi}=0 \nonu
\A \A {d^2 \zeta^1 \over d\phi^2}+\eta(r)\eta'(r) {dt \over
d\phi}{d \zeta^0 \over d\phi}-2r \eta(r) {d \zeta^3 \over
d\phi}\nonu
\A \A +\left[{1 \over 2}\left(\eta'^2(r)+\eta(r) \eta''(r)
\right)\left({dt \over d\phi}\right)^2-\left(\eta(r)+r\eta'(r)
\right)  \right]\zeta^1=0, \nonu
\A \A {d^2 \zeta^2 \over d\phi^2}+\zeta^2=0, \nonu
\A \A {d^2 \zeta^3 \over ds^2}+{2 \over r} {d \zeta^1 \over
d\phi}=0. \ea As is clear from the third equations of (33) that it
represent a simple harmonic motion, this means that the motion in
the plan $\theta=\pi/2$ is stable.

Assuming now the solution of the remaining equations given by \be
\zeta^0 = A_1 e^{i \omega \phi}, \qquad \zeta^1= A_2e^{i \omega
\phi}, \qquad \zeta^3 = A_3 e^{i \omega \phi},\ee where $A_1,
A_2,A_3$ are constants to be determined. From (34) and (33) we get
\ba \A \A {1 \over r
{r_1}^3\left({r_1}^3-({r_1}^3+3r^3)e^{-r^3/{r_1}^3} \right)}
\Biggl[{r_1}^6(r-6m)+(12m{r_1}^6-r{r_1}^6-9r^4{r_1}^3+42mr^3{r_1}^3
\nonu
\A \A  +9r^7-18mr^6)
e^{-r^3/{r_1}^3}-(6m{r_1}^6+42mr^3{r_1}^3+18mr^6)e^{-r^3/{r_1}^3}
\Biggr]>0, \ea which is the condition of the stability for the
vacuum non singular black hole solution.
\newsection{Main results}
The main results can be summarized  as follows \vspace{.3cm} \\ 1)
The singularity problem for the two solutions (22) and (23)
obtained before \cite{Ga} has been studied.\vspace{.3cm}\\ i) As
we see from (24) that all the scalars have a finite value as $r
\rightarrow 0$
 given by $96\displaystyle{m^2 \over {r_1}^6}$,
$144\displaystyle{m^2 \over {r_1}^6}$, $-24\displaystyle{m \over
{r_1}^3}$, $-12\displaystyle{m \over {r_1}^3}$,
$-18\displaystyle{m \over {r_1}^3}$, 0 and 0 respectively, i.e.,
Remain finite and tend to the de Sitter value for the scalars of
the Riemann
Christoffel tensor, Ricci tensor and Ricci scalar.\vspace{.3cm} \\
ii) For large r
 the scalars of (24) have the value $48\displaystyle{m^2 \over
r^6}$, 0, 0, $-9\displaystyle{m \over r^3}$, $-{9 \over 2}
\displaystyle{m \over r^3}$, $-{9 \over 2} \displaystyle{m \over
r^3}$ and 0 respectively which are the Schwarzschild value for the
Riemann Christoffel tensor, Ricci tensor and Ricci scalar.\vspace{.3cm}\\
iii) For the solution (25) we found that all the scalars have a
finite value as $r \rightarrow 0$
 given by $96\displaystyle{m^2 \over {r_1}^6}$,
$144\displaystyle{m^2 \over {r_1}^6}$, $-24\displaystyle{m \over
{r_1}^3}$, 0, 0, 0 and 0 respectively, i.e., Remain finite and
tend to the de Sitter value for the Riemann Christoffel tensor,
Ricci tensor and Ricci scalar same as of the solution (24). The
values of the scalars of the torsion, basic vector and traceless
part are different for the solutions (24) and (25) when $r
\rightarrow 0$. This may be due to the fact that the asymptotic
behavior of the parallel vector fields of the two solutions (24)
and (25)
is quite different.\vspace{.3cm}\\
iv) For large r
 the scalars of (24) have the value $48\displaystyle{m^2 \over
r^6}$, 0, 0,\\
$\displaystyle{2(4r^2-12rm-4r\sqrt{r^2-2mr}+8m\sqrt{r^2-2mr}+9m^2)
\over r^3(r-2m)}$,\\
$\displaystyle{(2r^2-4rm-2r\sqrt{r^2-2mr}+3m\sqrt{r^2-2mr})^2
\over r^4(r-2m)^2}$,
$\displaystyle{(r^2-2rm-r\sqrt{r^2-2mr}+3m\sqrt{r^2-2mr})^2 \over
r^4(r-2m)^2}$ and 0 respectively which are the Schwarzschild value
for the Riemann Christoffel tensor, Ricci tensor and Ricci scalar
same as of the solution (24). The values of the scalars of the
torsion, basic vector traceless are different for the solutions
(24) and (25) for large r due to
 the reason given above.

2) The stability condition for the metric of the vacuum non
singular black hole is
derived (35). From this condition we can see that.\vspace{.3cm}\\
i) As $r \rightarrow 0$ the value of (35) is
finite.\vspace{.3cm}\\ ii) As r becomes large  the value of (35)
takes the value $r>6m$ which the is condition of the stability of
the Schwarzschild solution.

\end{document}